\documentclass[10pt,twocolumn,letterpaper]{article}

\usepackage{iccv}
\usepackage{times}
\usepackage{epsfig}
\usepackage{graphicx}
\usepackage{amsmath}
\usepackage{amssymb}
\usepackage{subfigure}
\usepackage{xcolor}
\usepackage{caption2}
\usepackage{authblk}
\usepackage[pagebackref=true,breaklinks=true,letterpaper=true,colorlinks,bookmarks=false]{hyperref}

\iccvfinalcopy % *** Uncomment this line for the final submission

\begin{document}

%%%%%%%%% TITLE
\title{Non-local Attention Optimized Deep Image Compression}
\author[*]{Haojie Liu}
\author[*]{Tong Chen}
\author[*]{Peiyao Guo}
\author[*]{Qiu Shen}
\author[*]{Xun Cao}
\author[**]{Yao Wang}
\author[*]{Zhan Ma}
\affil[*]{Nanjing University}
\affil[**]{New York University}

%\author{Haojie Liu\\
%Nanjing Univeristy\\
%{\tt\small haojie@smail.nju.edu.cn}
%\and

\maketitle
%\thispagestyle{empty}

%%%%%%%%% ABSTRACT
\begin{abstract}
   %Popular learned image compression utilizes

This paper proposes a novel Non-Local Attention Optimized Deep Image Compression (NLAIC) framework, which is built on top of the  popular variational auto-encoder (VAE) structure.  Our NLAIC framework embeds non-local operations in the encoders and decoders for both image and latent feature probability  information (known as hyperprior) to capture both local and global correlations, and apply attention mechanism to generate  masks that are used to weigh the features for the image and hyperprior, which implicitly adapt bit allocation for different features based on their importance.   Furthermore,  both hyperpriors and spatial-channel neighbors of the latent features are used to improve entropy coding.  The proposed model outperforms
    the existing methods on Kodak dataset, including learned (e.g., Balle2019~\cite{minnen2018joint}, Balle2018~\cite{balle2018variational}) and conventional (e.g., BPG, JPEG2000, JPEG) image compression methods, for both PSNR and MS-SSIM distortion metrics.
\end{abstract}
%%%%%%%%% BODY TEXT

\section{Introduction}

Most recently proposed machine learning based image compression algorithms~\cite{balle2018variational,rippel2017real,mentzer2018conditional} leverage the autoencoder structure, which transforms raw pixels into compressible latent features via stacked convolutional neural networks (CNNs).  These latent features are entropy coded subsequently by exploiting the statistical redundancy. Recent prior works have revealed that compression efficiency can be improved when exploring the conditional probabilities via the contexts of spatial neighbors and hyperpriors~\cite{mentzer2018conditional,li2017learning,balle2018variational}. Typically, rate-distortion optimization~\cite{sullivan1998rate} is fulfilled by minimizing  Lagrangian cost $J$ = $R$ + $\lambda$$D$, when performing the end-to-end training. Here, $R$ is referred to as {\it entropy rate}, and $D$ is the {\it distortion} measured by either mean squared error (MSE) or multiscale structural similarity (MS-SSIM)~\cite{wang2003multiscale}.

\begin{figure}[t]
   \centering
   \includegraphics[scale=0.29]{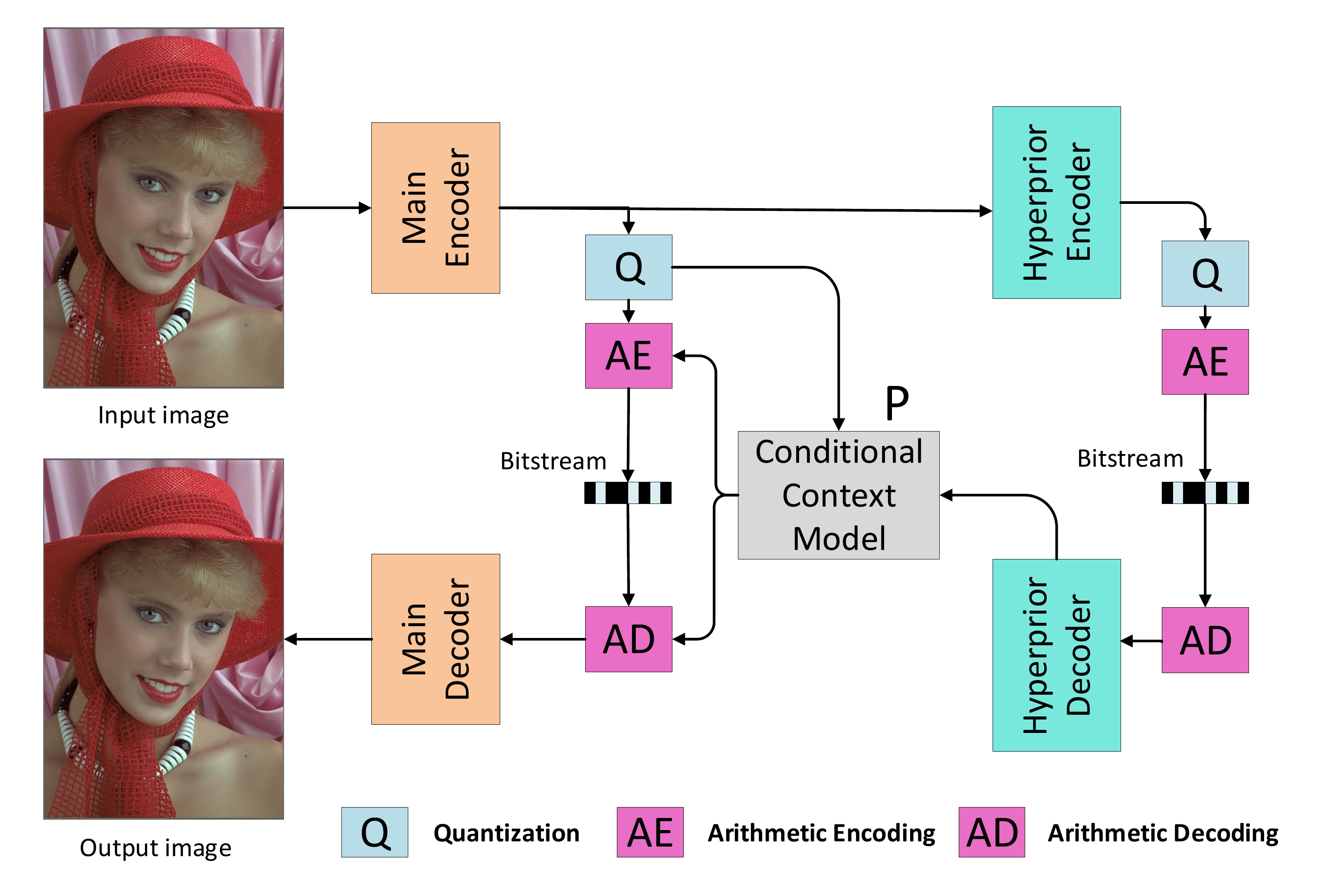}
   \caption{Proposed NLAIC framework using a variational autoencoder structure with embedded non-local attention optimization in the main and hyperprior encoders and decoders. }
   \label{framework}
\end{figure}

%Combined with entropy models, both autoregressive and hierarchical priors can be used for probabilistic prediction, yielding compressed bitstream with standard arithmetic coding. Traditional transforms with fixed basis such as discrete cosine transform (DCT) are limited to their transform capacity, which are replaced by convolution neural networks (CNNs) better, formulating an encoder between original signals and latent representation space of dimension reduction. Then a symmetric decoder,which plays a role of inverse function, maps these compressible features into pixel domain. Estimating the entropy represented by the potential features precisely with different conjunct entropy models~\cite{mentzer2018conditional,li2017learning,balle2018variational} can further improve the compression performance. In general, a rate-distortion optimization (RDO) problem during training is the key to achieve the least distortion for a given constraint,i.e., $J$ = $R$ + $\lambda$$D$, with $R$ for the rate and $D$ for the distortion such as mean square error(MSE) or MS-SSIM.
However, existing methods still present several limitations. For example, most of the operations, such as stacked convolutions, are performed locally with limited receptive field, even with pyramidal decomposition. Furthermore, latent features are treated with equal importance in either spatial or channel dimension in most works, without considering the diverse visual sensitivities to  various contents (such as texture and edge). Thus, attempts have been made in~\cite{li2017learning,mentzer2018conditional} to exploit importance maps on top of  latent feature vectors for adaptive bit allocation. But these methods require the extra explicit signaling overhead to carry the importance maps.

%However, current compression methods have several limitations mainly due to the relative smaller receptive field with stacked CNNs. Although effective transforms (e.g.,generalized divisive normalization (GDN))~\cite{balle2016end} or novel feature extraction method~\cite{rippel2017real} (e.g.,pyramidal decomposition) are proposed and proven to be benefit to the whole coding framework, it is difficult for information flow and propagation as larger distant positions in the neural network. In addition, most of the existing coding algorithms treat the features equally in the spatial and feature domain, and ignore the diversity of texture and shape. For this reason, \cite{li2017learning,mentzer2018conditional} proposed importance map as a space-channel wise mask and allocate the bits for different regions. The mask generation is simple and element-wise multiplicated with the quantized features, resulting in a problem if encoding a mask at different bit rate. The performance of the overall framework is far from the best.

In this paper, we introduce {\it non-local operation} blocks proposed in~\cite{wang2018non} into the variational autoencoder (VAE) structure to capture both local and global correlations among pixels, and generate the attention masks which help to yield more compact distributions of  latent features and hyperpriors.% Additionally, we also try to explore the non-uniform importance of the content for better compression.
Different from those existing methods in~\cite{li2017learning,mentzer2018conditional}, we use non-local processing to generate attention masks at  different layers (not only for quantized features), to allocate the bits intelligently through the end-to-end training. We also improve  the context modeling of the entropy engine for better latent feature compression, by using a masked 3D CNN (i.e., 5$\times$5$\times$5) on the latent features to generate the conditional statistics of the latent features.

Two different model implementations are provided, one is the ``NLAIC joint'' , which uses both hyperpriors and spatial-channel neighbors in the latent features for context modeling, and the other  is the ``NLAIC baseline'' with contexts only from hyperpriors. Our joint model outperforms all existing learned and traditional image compression methods, in terms of the rate distortion efficiency for the distortion measured by both MS-SSIM and PSNR.
%To the best of our knowledge, our baseline model is the first one demonstrating the better PSNR over BPG~\cite{BPG}     at the same bit rate with significant computation reduction.

%non-local quantized latent features and non-local hierarchical priors both. Assume that the previous methods only depending on the local convolutions, non-local priors can provide more accurate probability prediction, benefit from extra global information. Meanwhile, quantized errors usually lead to varying degrees of recovery problems in the plain and complicated regions which are restored difficultly on details abound. What is different from \cite{li2017learning,mentzer2018conditional} is that we add attention mechanism at different scales and combine it with non-local blocks to make the network pay more attention to the challenging areas. Intuitively,it is also useful extend the probability eatimation with autoregressive information (e.g.PixelCNN and PixelRNN~\cite{oord2016pixel}) for performance improvement. Here, we introduce a single 5$\times$5$\times$5 3D linear convolution in conjunction with VAE and find it more than 5\% bitrate reduction with jointly training.

To further verify the efficiency of our framework, we also conduct ablation studies to discuss model variants such as removing non-local operations and attention mechanisms layer by layer, as well as the visual comparison. These additional experiments provide further evidence of the superior performance of our proposed NLAIC framework over a broad dataset.

The main contributions of this paper are highlighted as follows:
\begin{itemize}
   \item We are the {\it first} to introduce {\it non-local operations} into compression framework to capture both local and global correlations among the pixels in the original image and feature maps.%to generate the attention maps which is benefit to obtain more compact latent features.
   \item We apply {\it attention mechanism} together with aforementioned non-local operations to generate implicit {\it importance masks} to guide the adaptive processing of latent features. These masks essentially allocate more bits to more important features that are critical for reducing the image distortion.% by allocating more information to challenging area (e.g., more bits to salient region, and vice versa);
   \item We employ a one-layer masked 3D CNN to exploit the spatial and cross channel correlations in the latent features, the output of which is then concatenated with  hyperpriors to estimate the conditional statistics of the latent features, enabling more efficient entropy coding.
   %We utilize a one-layer masked 3D CNN-based structure to leverage autoregressive neighbors, concatenate it with the reconstructed features from hyperpriors  and fuse them with 1$\times$1$\times$1 convolutions for better context modeling of entropy coding.
   %not only for the latent quantized features but also for the hierarchical priors, removing more redundancies. We find it already outperform the current leading methods including~\cite{minnen2018joint} measured by MS-SSIM, and outperform BPG measured by PSNR without extra context modeling.
\end{itemize}
\begin{table*}[t]\footnotesize
   \centering
   \caption{ Detailed Parameter Settings in NLAIC as shown in Fig.~\ref{framework}: ``Conv" denotes a convolution layer with kernel size and number of output channels. ``s" is the stride (e.g.,s2 means a down/up-sampling with stride 2). NLAM represents the non-local attention modules. ``$\times$3" means cascading 3 residual blocks (ResBlock).}
   \label{tab:parameters}
   \begin{tabular}{|c|c|c|c|c|}

     \hline
   Main Encoder & Main Decoder & Hyperprior Encoder &  Hyperprior Decoder & Conditional Context Model\\
     \hline
     Conv: 5$\times$5$\times$192 s2 & NLAM &  ResBlock($\times$3): 3$\times$3$\times$192 & NLAM & Masked: 5$\times$5$\times$5$\times$24 s1\\
     ResBlock($\times$3): 3$\times$3$\times$192 & Deconv: 5$\times$5$\times$192 s2 &  Conv: 5$\times$5$\times$192 s2 & Deconv: 5$\times$5$\times$192 s2 & Conv: 1$\times$1$\times$1$\times$48 s1\\
     Conv: 5$\times$5$\times$192 s2 &ResBlock($\times$3): 3$\times$3$\times$192 & ResBlock($\times$3): 3$\times$3$\times$192 & ResBlock($\times$3): 3$\times$3$\times$192 & ReLU\\
     NLAM & Deconv: 5$\times$5$\times$192 s2 & Conv: 5$\times$5$\times$192 s2 &Deconv: 5$\times$5$\times$192 s2 & Conv: 1$\times$1$\times$1$\times$96 s1\\
     Conv: 5$\times$5$\times$192 s2 & NLAM & NLAM & ResBlock($\times$3): 3$\times$3$\times$192 & ReLU\\
     ResBlock($\times$3): 3$\times$3$\times$192& Deconv: 5$\times$5$\times$192 s2 & &Conv: 5$\times$5$\times$384 s1 & Conv: 1$\times$1$\times$1$\times$2 s1\\
     Conv: 5$\times$5$\times$192 s2 & ResBlock($\times$3): 3$\times$3$\times$192&&&\\
     NLAM & Conv: 5$\times$5$\times$3 s2&&&\\
     \hline
   \end{tabular}
 \end{table*}

%-------------------------------------------------------------------------
\section{Related Work} \label{sec:related_work}
\textbf {Non-local Operations.} Most traditional filters (such as Gaussian and mean)                process the data locally,
by using a weighted average of spatially neighboring pixels. It usually produces over-smoothed reconstructions.
Classical non-local methods for image restoration problems (e.g., low-rank modeling~\cite{gu2014weighted}, joint sparsity~\cite{mairal2009non} and non-local means~\cite{buades2005non}) have shown their superior efficiency for quality improvement by exploiting non-local correlations. Recently, non-local operations haven been included into the deep neural networks (DNN) for video classification~\cite{wang2018non}, image restoration (e.g., denoising, artifacts removal and super-resolution)~\cite{liu2018non, zhang2018residual}, etc, with significant performance improvement reported. It is also worth to point out that non-local operations have been applied in other scenarios, such as intra block copy in screen content extension of the High-Efficiency Video Coding (HEVC)~\cite{IBC}.

%designed a non-local module to produce feature correlations for self-similarity,formulating a non-local recurrent network (NLRN) for image restoration. \cite{zhang2018residual} was also inspired by this and combined non-local modules with attention, leading the performance on image denosing, artifacts reduction and super-resolution. In traditional compression methods, block matching or intra block copy also has the idea of non-local for some special scenarios.

\textbf {Self Attention.}
Self-attention mechanism is widely used in deep learning based natural language processing (NLP)~\cite{luong2015effective,firat2016multi,vaswani2017attention}. It can be described as a mapping strategy which  queries a set of key-value pairs to an output. For example, Vaswani {\it et. al} \cite{vaswani2017attention} have proposed multi-headed attention methods which are extensively used for machine translation. For those low-level vision tasks~\cite{zhang2018residual,li2017learning,mentzer2018conditional}, self-attention mechanism makes generated features with spatial adaptive activation and enables adaptive information allocation with the emphasis on more challenging areas (i.e., rich textures, saliency, etc).

In image compression, quantized attention masks are commonly used for adaptive bit allocation, e.g., Li {\it et. al}~\cite{li2017learning} uses 3 layers of local convolutions and Mentzer {\it et. al}~\cite{mentzer2018conditional} selects one of the quantized features.
Unfortunately, these  methods require the extra explicit signaling overhead.
Our model adopts  attention mechanism that is close to~\cite{li2017learning,mentzer2018conditional} but
applies multiple layers of non-local as well as convolutional operations to automatically generate attention masks from the input image. The attention masks  are applied to the temporary latent features directly to generate the final latent features to be coded.
Thus, there is no need to use extra bits to code the masks.

%In addition, we put attention mechanism at different scales, however, \cite{li2017learning,mentzer2018conditional} only consider it at the end of the encoder. Inspired by~\cite{zhang2018residual}, we further introduce non-local attention operations for hierarchical priors, and experiments are shown that our non-local priors can help remove more structural redundancies, estimating the entropy accurately.

\textbf {Image Compression Architectures.} DNN based image compression generally relies on well-known autoencoders.
Its back propagation scheme requires all the steps differentiable in an end-to-end manner. Several methods (e.g., adding uniform noise~\cite{balle2016end}, replacing the direct derivative with the derivative of the expectation~\cite{TodericiVJHMSC16} and soft-to-hard quantization~\cite{agustsson2017soft}) are developed to approximate the non-differentiable quantization process.
On the other hand, entropy rate modeling of quantized latent features is another critical issue for learned image compression. PixelCNNs~\cite{oord2016pixel} and VAE are commonly used for entropy estimation following the Bayesian generative rules.
Recently, conditional probability estimates based on autoregressive neighbors of the latent feature maps and hyperpriors jointly has shown significant improvement in entropy coding.

\section{Non-Local Attention Implementation}

\subsection{General Framework}
\begin{figure*}[t]
   \centering
   \subfigure[]{\includegraphics[scale=0.32]{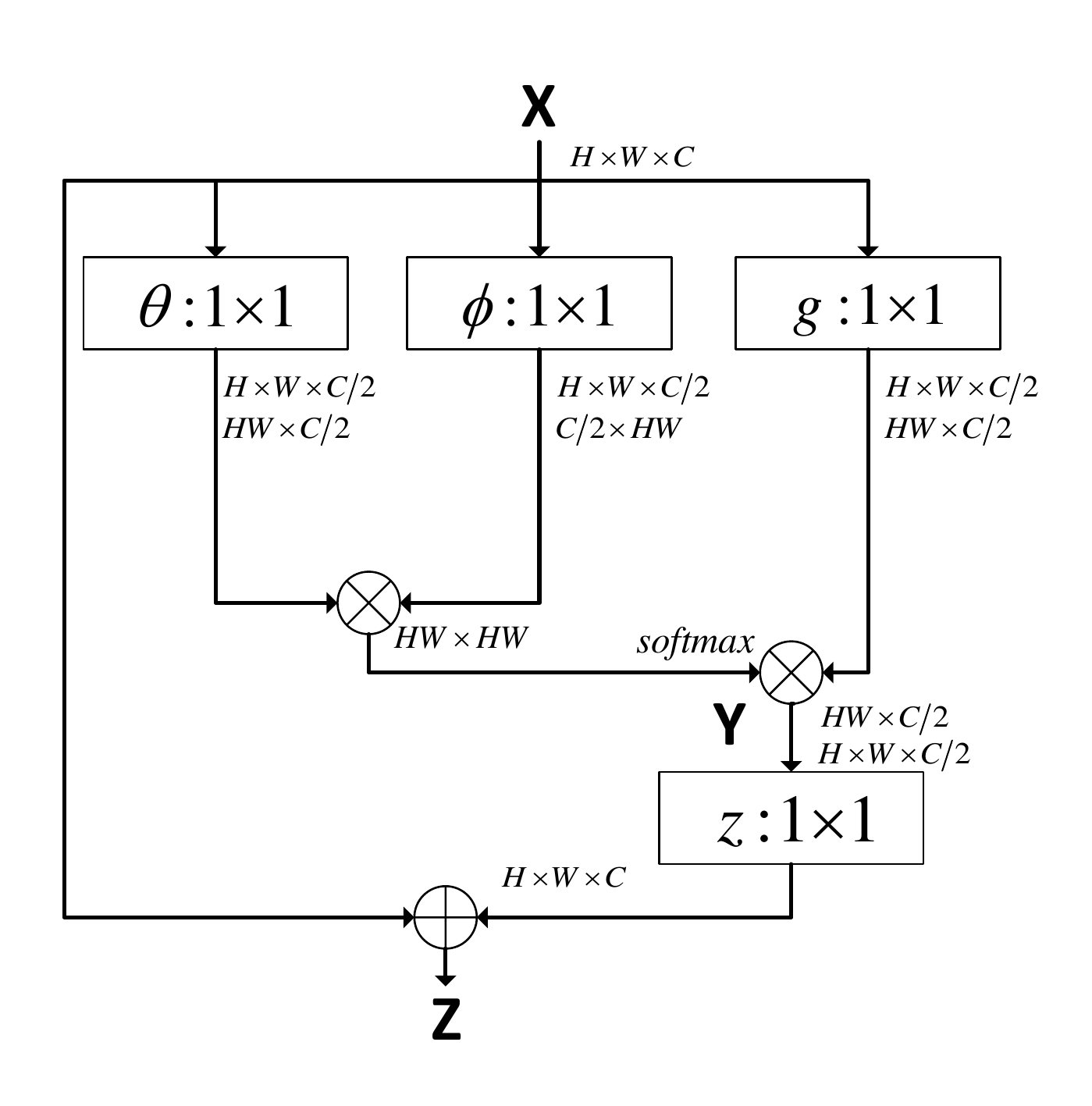} \label{sfig:non_local_fig}}
   \subfigure[]{\includegraphics[scale=0.32]{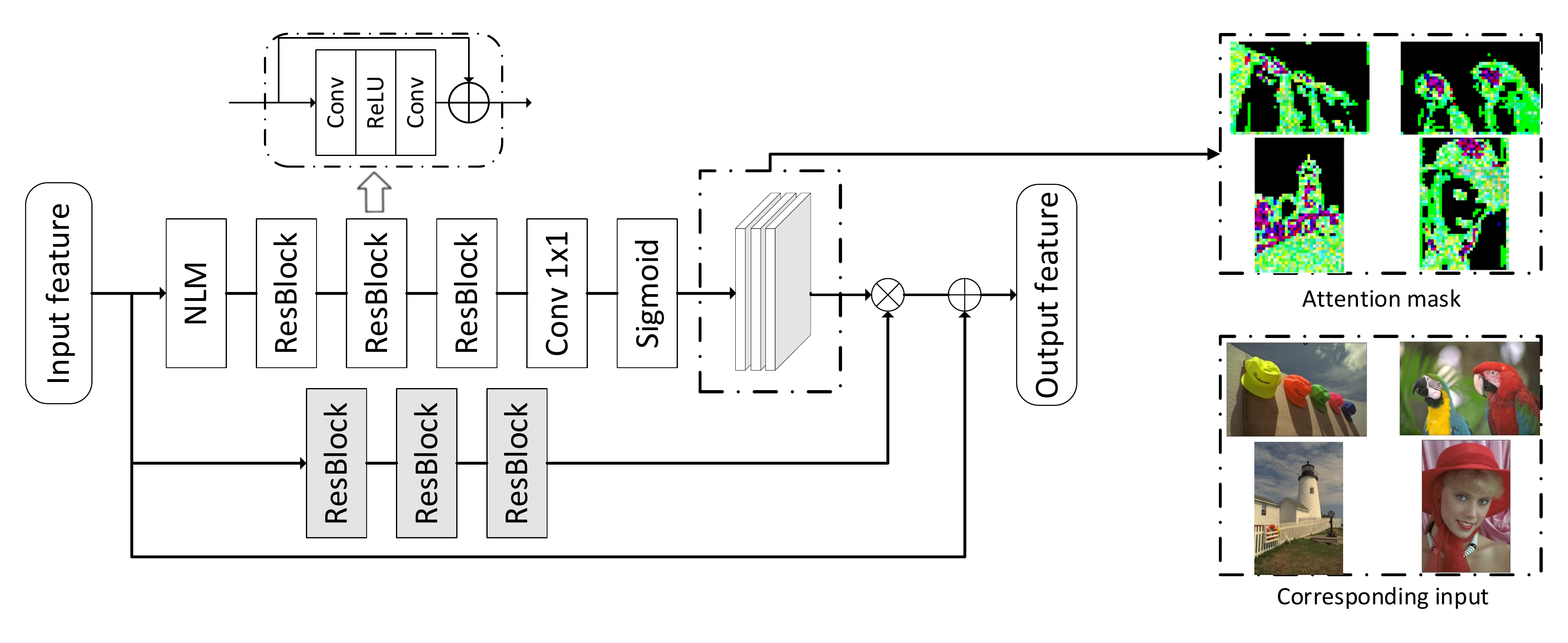}\label{sfig:non_local_attention}}
   \caption{(a) Non-local module (NLM). $H\times{W}\times{C}$ denotes the size of fearure maps with height $H$, width $W$ and channel $C$. $\oplus$ is the add operation and $\otimes$ is the matrix multiplication. (b) Non-local attention module (NLAM). The main branch consists of 3 residual blocks.The mask branch combines non-local modules with residual blocks for attention mask generation. The details of residual blocks are shown in the dash frame.}
   \label{non_local_attention_fig}
\end{figure*}
Fig.~\ref{framework} illustrates our NLAIC framework. It is built on a variational autoencoder structure~\cite{balle2018variational}, with {\it non-local attention modules} (NLAM) as basic units in both main and hyperprior encoder-decoder pairs (i.e., ${\mathbb E}_M$, ${\mathbb D}_M$, ${\mathbb E}_h$ and ${\mathbb D}_h$). ${\mathbb E}_M$ with quantization $Q$ are used to generate latent quantized features and ${\mathbb E}_D$ decodes the features into the reconstructed image. ${\mathbb E}_h$ and ${\mathbb D}_h$ generate much smaller side information as hyperpriors.
The hyperpriors as well as autoregressive neighbors of the latent features are then processed through the conditional context model ${\mathbb P}$ to generate the conditional probability estimates for entropy coding of the latent quantized features.

 %for improving the conditional probability modeling $\mathbb P$, in addition to those spatial-channel neighbors. %Specifically, A masked 5$\times$5$\times$5 3D CNN-based prediction is first utilized to leverage the contexts from autoregressive neighbor pixels, and then followed by 1$\times$1$\times$1 convoluted channel information after concatenating hyperpriors.
Table~\ref{tab:parameters} details the network structures and associated parameters of five different components in the proposed NLAIC framework.  The NLAM module is shown in Fig.~\ref{sfig:non_local_attention}, and explained in Sections~\ref{sec:NLM} and~\ref{sec:NLAM}.

%Parameter details are listed in . Note that in Table~\ref{tab:parameters}, we use 3D convolution layers with weights sharing in channel dimension. We use a 2D convolution to transform the decoded hyper features into the same shape as latent quantized features, and concatenate it with the masked features, feeding into the following 1$\times$1$\times$1 convolutions.

\subsection{Non-local Module}\label{sec:NLM}

Our NLAM adopts the non-local network proposed in ~\cite{wang2018non} as a basic block, as shown in Fig.~\ref{non_local_attention_fig}. As shown in Fig.~\ref{sfig:non_local_fig}, the non-local module (NLM) computes the output at pixel $i$, $Y_i$, using a weighted average of the transformed feature values at pixel $j$, $X_j$, as below:
\begin{equation}
Y_i = \frac{1}{C(X)}{\sum_{{\forall}j}}f(X_i,X_j)g(X_j),
\label{Eq1}
\end{equation}
where $i$ is the location index of output vector $Y$ and $j$ represents the index that enumerates all accessible positions of input $X$. $X$ and $Y$ share the same size. The function $f(\cdot)$ computes the correlations between $X_i$ and $X_j$, and  $g(\cdot)$ computes the representation of the input at the position $j$. $C(X)$ is a normalizing factor to generate the final response which is set as $C(X) = \sum_{\forall j}f(X_i,X_j)$. Note that a variety of function forms of $f(\cdot)$ have been already discussed in~\cite{wang2018non}. Thus in this work, we directly use the embedded Gaussian function for $f(\cdot)$, i.e.,
%Since ~\cite{wang2018non} has discussed different functions of $f(.)$ for evaluating the relationship between $X_i$ and $X_j$, we do not focus on this.
%As shown in Figure~\ref{non_local_fig}, we choose embedded Gaussian fuction for $f(.)$ which is described as:
\begin{equation}
f(X_i,X_j) = e^{\theta(X_i^T)\phi(X_j)}.
\label{Eq2}
\end{equation}
Here, $\theta(X_i) = W_{\theta}{X_i}$ and $\phi(X_j) = W_{\phi}{X_j}$, where $W_{\theta}$ and $W_{\phi}$ denote the cross-channel transform using  1$\times$1 convolution  in our framework. The weights $f(X_i,X_j)$ are further modified by a softmax operation. The operation defined in Eq.~\eqref{Eq1} can be written in matrix form~\cite{wang2018non} as:
\begin{equation}
Y = {\tt softmax}({X}^TW_{\theta}^TW_{\phi}{X})g(X). \label{eq:nl_Y}
\end{equation}
In addition, residual connection can be applied for better convergence as suggested in~\cite{wang2018non}, as shown in Fig.~\ref{sfig:non_local_fig}, i.e.,
\begin{equation}
Z_i = W_{z}Y_i+X_i, \label{eq:nl_Z}
\end{equation}
where $W_z$ is also a linear 1$\times$1 convolution across all channels, and $Z_i$ is the final output vector.
%-------------------------------------------------------------------------
\subsection{Non-local Attention Module}\label{sec:NLAM}
Importance map has been adopted in~\cite{li2017learning,mentzer2018conditional} to adaptively allocate information to quantized latent features. For instance, we can give more bits to textured area but less bits to elsewhere, resulting in better visual quality at the similar bit rate. Such adaptive allocation can be implemented by using an explicit {\it mask}, which must be specified with additional bits. As aforementioned, existing mask generation methods in~\cite{li2017learning,mentzer2018conditional} are too simple to handle areas with more complex content characteristics.

%introduce importance maps for quantized latent features. They mask unimportant positions of features and allocate more information at challenging regions, which achieves better visual quality compared with these balance approches. However, the weighted mask directly affects the size of bitstream and sometimes cost a little bits. Moreover, the importance map generation is usually too simple and cannot handle more complicated cases.

Inspired by~\cite{zhang2018residual}, we propose to use a cascade of a non-local module and  regular convolutional layers to generate the attention masks, as shown in
Fig.~\ref{sfig:non_local_attention}. The NLAM consists of two branches. The main branch uses conventional stacked networks to generate features and the  mask branch  applies the NLM with three residual blocks~\cite{he2016deep}, one 1$\times$1 convolution and {\tt sigmoid} activation to produce a joint {\it spatial-channel attention mask} $M$, i.e.,\begin{equation}
   M = {\tt sigmoid}(F_{\rm NLM}(X)), \label{eq:attention_mask}
   \end{equation}
   where $M$ denotes the attention mask and $X$ is the input features. $F_{\rm NLM}(\cdot)$ represents the operations of using NLM with subsequent three residual blocks and 1$\times$1 convolution which are shown in Fig.~\ref{sfig:non_local_attention}.
This attention mask $M$, having its element $0<M_k<1, M_k\in{\mathbb{R}}$, is element-wise multiplied with feature maps from the main branch to perform adaptive processing.
Finally a residual connection is added for faster convergence.
 %The values of attention mask range from 0 to 1 continuously.It can be described as:

We avoid any batch normalization (BN) layers and only use one ReLU in our residual blocks, justified through our experimental observations.

Note that in existing learned image compression methods, particularly for those with
superior performance~\cite{balle2016end,balle2018variational,minnen2018joint,liu2019gated}, GDN activation has
proven its better efficiency compared with ReLU, tanh,  sigmoid, leakyReLU, etc.  This may be due to the fact that
 GDN captures the global information across all feature channels  at the same pixel location. However, we just use the simple ReLU function, and rely on our proposed NLAM to capture both the local and global correlations.
We also find through experiments that inserting two pairs of two layers of NLAM for the main encoder-decoder, and one layer of NLAM in the hyperprior encoder-decoder, provides the best performance.
As will be shown in subsequent Section~\ref{sec:experiment}, our NLAIC has demonstrated the state-of-the-art coding efficiency.

%A series of proposed works~\cite{balle2016end,balle2018variational,minnen2018joint} are based on GDN, which is proven to be more effective than common activation functions (e.g.,ReLU, tanh, sigmoid and leakyReLU), leading to the best performance. They heavily depend on local CNNs and ignore non-local information and spatial differences. In our NLAIC, we introduce NLAM not only for main encoder and decoder but also for the hierarchical encoder and decoder to solve these problems. Just as we expected, NLAM makes main encoder-decoder network for better reoconstruction, and is more effecient for hierarchical encoder-decoder to estimate the entropy. It mainly profits from the global information extraction and attention mechanism which can predict the complicated region accurately even without GDN. More experiments such as removing NLAM are conducted to make ablation study in the following section.
\begin{figure}[b]
   \centering
   \includegraphics[scale=0.31]{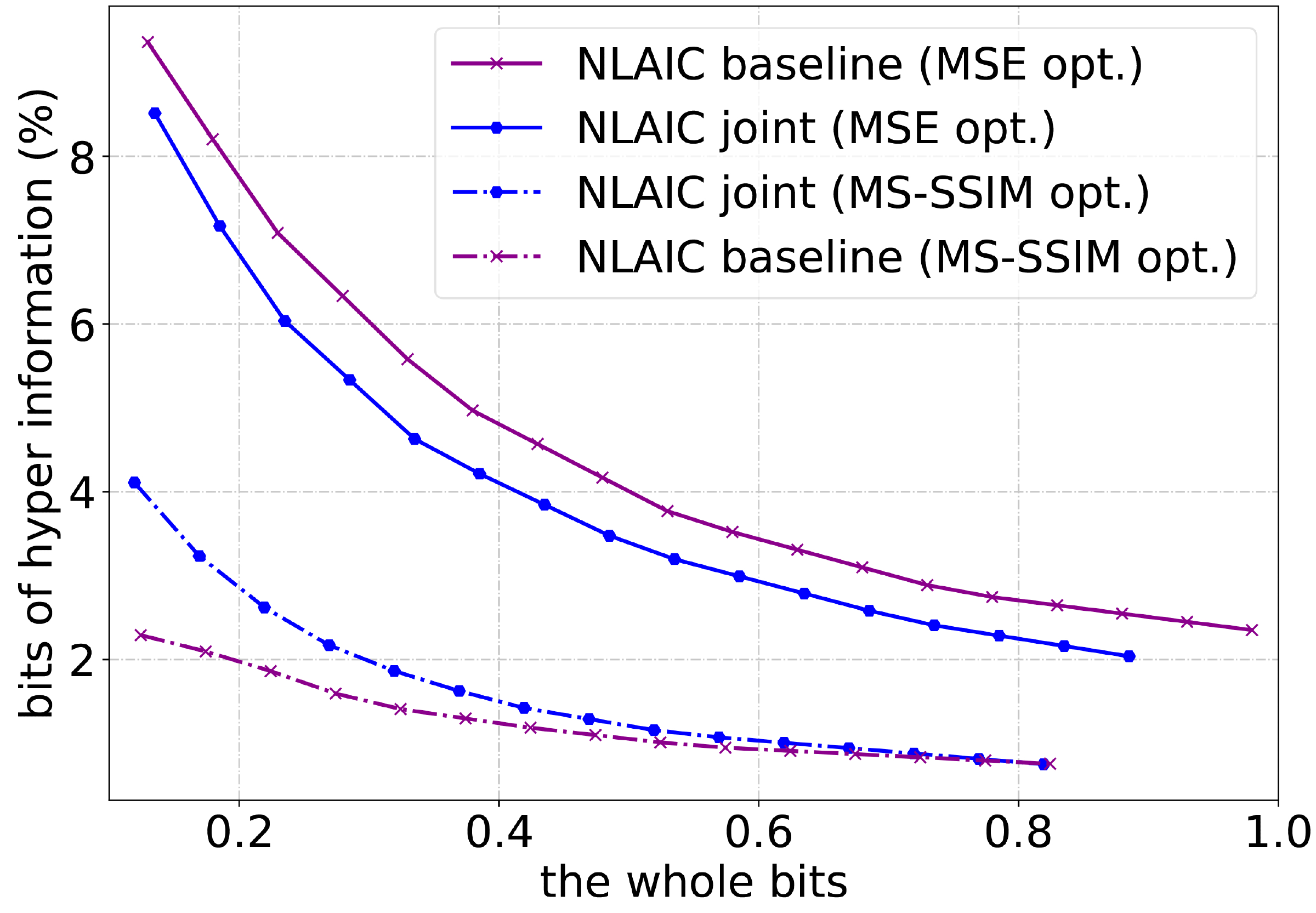}
   \caption{Illustration of percentage of $\hat{z}$ in the entire bitstream. For the case that model is optimized using MSE loss, $\hat{z}$ occupies less percentage for joint model than the baseline; But the outcome is reversed for the case that model is tuned with MS-SSIM loss. The percentage of $\hat{z}$ for MSE loss optimized method is noticeably higher than the scenario using MS-SSIM loss.}
   \label{hyper_bits_comsuming}
\end{figure}
\subsection{Entropy Rate Modeling}
Previous sections present our novel NLAM scheme to transform the input pixels into more compact latent features.
This section details the entropy rate modeling part that is critical for the overall rate-distortion efficiency.
\begin{figure*}[t]
\centering
\subfigure[]{\includegraphics[scale=0.32]{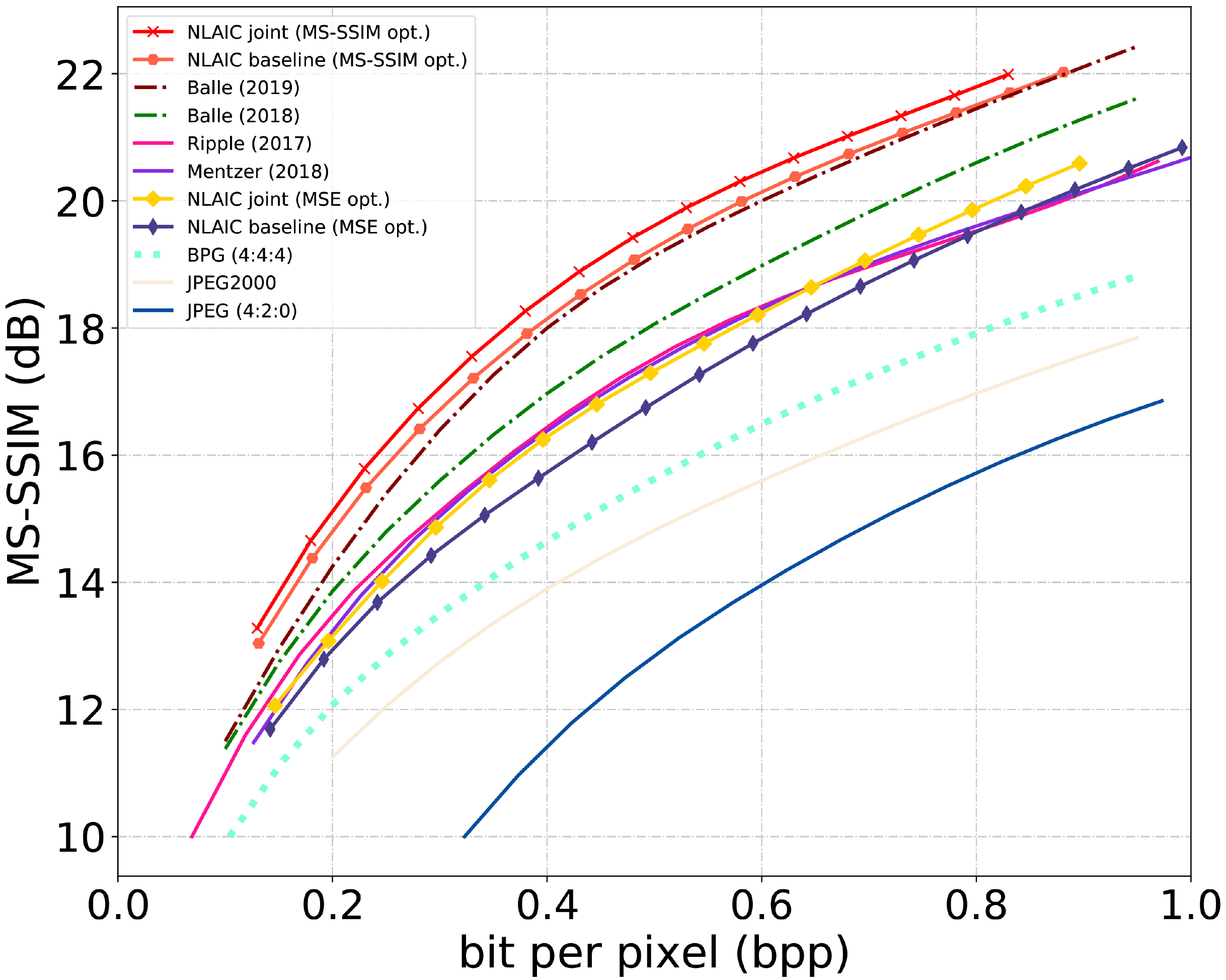} \label{sfig:ssim_perf}}
\subfigure[]{\includegraphics[scale=0.32]{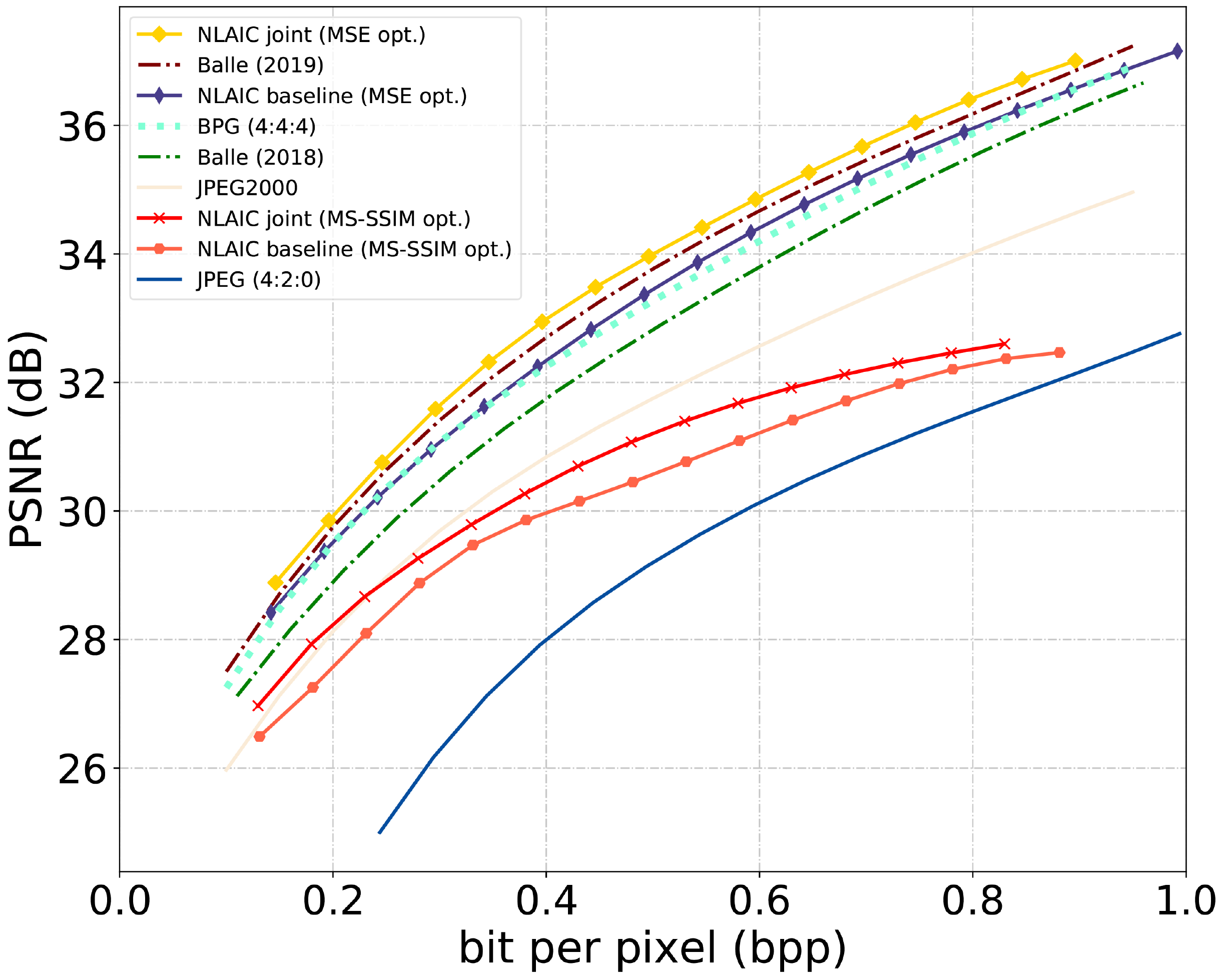} {\label{sfig:psnr_perf}}}

\caption{Illustration of the rate-distortion performance on Kodak. (a) distortion is measured by MS-SSIM (dB). Here we use $-10\log_{10}(1-d)$ to represent raw MS-SSIM ($d$) in dB scale. (b)  PSNR is used for distortion measurement.
}
\label{rd_curve}
\end{figure*}
\subsubsection{Context Modeling Using Hyperpriors} \label{sssec:context_z}
Similar as~\cite{balle2018variational}, a non-parametric, fully factorized density model is used for hyperpriors $\hat{z}$, which is described as:
\begin{equation}
  p_{\hat{z}|\psi}(\hat{z}|\psi) = {\prod_i} ( p_{z_i|\psi^{(i)}}(\psi^{(i)})*\mathcal{U}(-\frac{1}{2},\frac{1}{2})) (\hat{z}_i), \label{eq:prob_z}
 \end{equation}
where $\psi^{(i)}$ represents the parameters of each univariate distribution $p_{\hat{z}|\psi^{(i)}}$.

For quantized latent features $\hat{y}$, each element $\hat{y}_i$ can be modeled as a conditional Gaussian distribution as:
\begin{equation}
  p_{\hat{y}|\hat{z}}(\hat{y}|\hat{z}) = {\prod_i} (\mathcal{N}(\mu_i,{\sigma_i}^2) *\mathcal{U}(-\frac{1}{2},\frac{1}{2})) (\hat{y}_i), \label{eq:prob_y_z}
 \end{equation}
where its $\mu_i$ and $\sigma_i$ are predicted using the distribution of $\hat{z}$. We evaluate the bits of $\hat{y}$ and $\hat{z}$ using:
\begin{align}
  R_{\hat{y}} &= - {\sum\nolimits_i} {\log_2}(p_{\hat{y}_i|\hat{z}_i}(\hat{y}_i|\hat{z}_i)), \label{eq:rate_y}\\
 R_{\hat{z}} &= -{\sum\nolimits_i}  {\log_2}(p_{\hat{z}_i|\psi^{(i)}}({\hat{z}_i}|\psi^{(i)})). \label{eq:rate_z}
\end{align}
Usually, we take $\hat{z}$ as side information for estimating $\mu_i$ and $\sigma_i$  and $\hat{z}$ only occupies a very small fraction of bits, shown in Fig.~\ref{hyper_bits_comsuming}. %And we find our NLAIC shows its superiority such as precise prediction with approximate bits of $\hat{z}$, although $\hat{y}$ even contains more information, helping reconstruct higher quality image.

\subsubsection{Context Modeling Using Neighbors} \label{sssec:context_joint}
PixelCNNs and PixelRNNs~\cite{oord2016pixel} have been  proposed for effective modeling of probabilistic distribution of images using local neighbors in an autoregressive way. It is further extended for adaptive context modeling in compression framework with noticeable improvement. For example, Minnen {\it et al.}~\cite{minnen2018joint} have proposed to extract autoregressive information by a 2D 5$\times$5 masked convolution, which is combined with hyperpriors using stacked 1$\times$1 convolution, for  probability estimation. It is the {\it first} deep-learning based method with better PSNR compared with the BPG444  at the same bit rate.

\begin{figure}[b]
   \centering
   \includegraphics[scale=0.5]{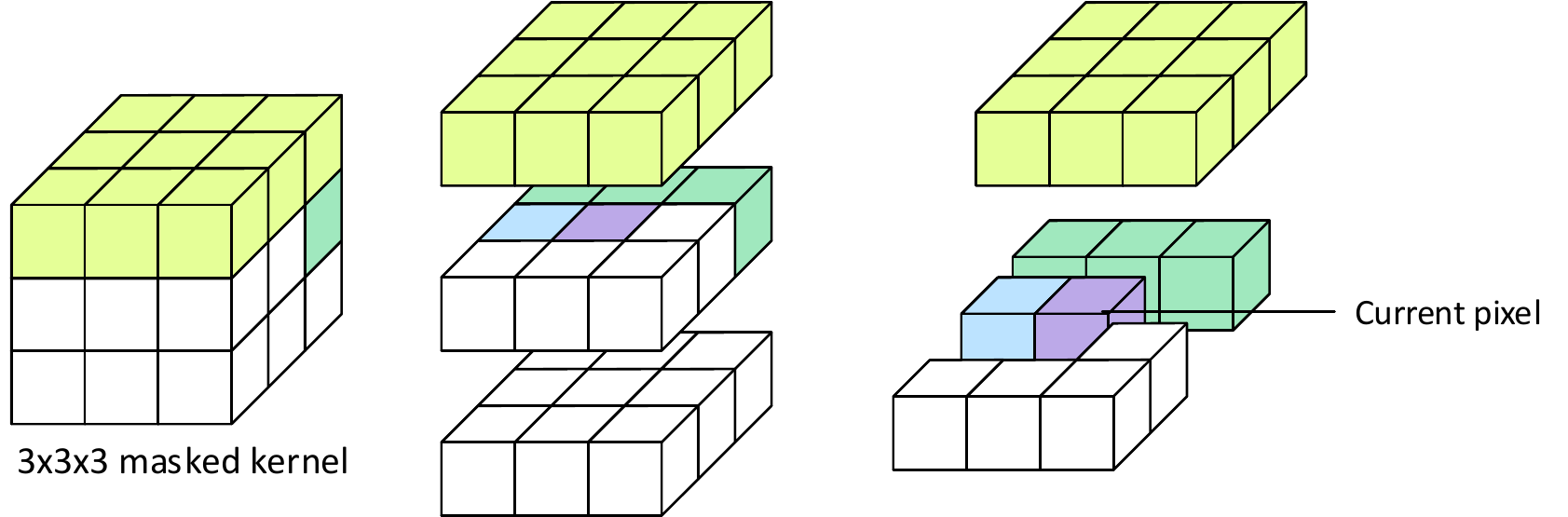}
   \caption{In 3$\times$3$\times$3 masked convolution, the current pixel  (in purple) is predicted by  the processed pixels (in yellow, green and blue) in a 3D space. The unprocessed pixels (in white) and the current pixel are masked with zeros.}
   \label{maskconv3d}
\end{figure}

In our NLAIC, we use a one-layer 5$\times$5$\times$5
3D masked convolution to exploit the spatial and cross-channel correlation. For simplicity, a  3$\times$3$\times$3 example is
shown in Fig.~\ref{maskconv3d}.
 Traditional 2D PixelCNNs need to search for a well structured channel order to exploit the conditional probability efficiently. Instead, our proposed 3D masked convolutions implicitly exploit correlation among adjacent channels.
 Compared to 2D masked CNN used in  ~\cite{minnen2018joint}, our 3D CNN approach significantly reduces the network parameters for the conditional context modeling.
% Together with the weights sharing in channel dimension, our 3D masked convolutions  consumes a fairly small-size network parameters.
Leveraging the additional contexts from neighbors via an autoregressive fashion, we can %significantly reduce the entropy (hence the rate) of the latent features.
obtain a better conditional Gaussian distribution to model the entropy as:
\begin{align}
  p_{\hat{y}}(\hat{y}_i|\hat{y}_1,..., &\hat{y}_{i-1},\hat{z})  = \nonumber\\
   & {\prod_i} (\mathcal{N}(\mu_i,{\sigma_i}^2) *\mathcal{U}(-\frac{1}{2},\frac{1}{2})) (\hat{y}_i), \label{eq:context_joint}
 \end{align}
 where  $\hat{y}_1, {\hat {y}}_2,..., \hat{y}_{i-1}$ denote the causal (and possibly reconstructed) pixels prior to current pixel $\hat{y}_i$. %Obtaining autoregresive information, the cost of $\hat{z}$ fluctuates slightly at different bit rate which is shown in Figure~\ref{hyper_bits_comsuming}.

  %In the normal sense, addtional context model helps predict latent entropy losslessly and we will cost less $\hat{z}$. Our results optimized for PSNR follow this rule. However, it is surprising that the percentage of $\hat{z}$ almostly remains unchanged in case of optimizing for MS-SSIM, and addtional context modeling even makes it cost more in practice. We think it has a certain relationship with MS-SSIM. The left graph of Figure~\ref{rd_curve} shows that optimized for MS-SSIM makes PSNR grow very slowly but we model the entropy with conditional Gaussian distribution which is close to MSE in actual.

%------------------------------------------------------------------------
\section{Experiments}\label{sec:experiment}

\subsection{Training}
 We use \textbf{COCO}~\cite{lin2014microsoft} and \textbf{CLIC}~\cite{clic}  datasets to train our NLAIC framework. We randomly crop images into  192$\times$192$\times$3 patches for subsequent learning. Rate-distortion optimization (RDO) is applied to do end-to-end training at various bit rate, i.e.,
\begin{align}
L = {\lambda}{\cdot}d(\hat{x},x)+R_y+R_z. \label{eq:rdo}
\end{align}
$d(\cdot)$ is a distortion measurement between reconstructed image $\hat{x}$ and the original image $x$. Both negative MS-SSIM and MSE are used in our work as distortion loss for evaluation, which are marked as ``MS-SSIM opt.'' and ``MSE opt.'', respectively.
$R_y$ and $R_z$ represent the estimated bit rates of latent features and hyperpriors, respectively.
Note that all components of our NLAIC are trained together.
We set learning rates (LR) for ${\mathbb E}_M$, ${\mathbb D}_M$, ${\mathbb E}_h$, ${\mathbb D}_h$ and
${\mathbb P}$ at $3 {\times} 10^{-5}$ in the beginning.
But for ${\mathbb P}$, its LR  is clipped to  $10^{-5}$  after 30 epochs.
Batch size is set to 16 and the entire model is trained on 4-GPUs in parallel.

To understand the contribution of the context modeling using spatial-channel neighbors, we offer two different implementations: one is ``NLAIC baseline'' that only uses the hyperpriors to estimate the means and variances of the latent features (see Eq.~\eqref{eq:prob_y_z}), while the other is ``NLAIC joint'' that uses both hyperpriors and previously coded pixels in the latent feature maps (see Eq.~\eqref{eq:context_joint}). In this work, we {\it first} train the ``NLAIC baseline'' models. To train the
``NLAIC joint" model,
one way is fixing the main and hyperprior encoders and decoders in the baseline model, and updating  only the conditional context model $\mathbb P$.  Compared with the ``NLAIC baseline'', such transfer learning based ``NLAIC joint'' provides 3\% bit rate reduction at the same distortion. Alternatively, we could use the baseline models as the start point, and refine all the modules in the ``NLAIC joint'' system. In this way,   ``NLAIC joint'' offers more than 9\% bit rate reduction over the ``NLAIC baseline'' at the same quality. Thus, we choose the latter one for better performance.
\begin{figure}[t]
   \centering
   \includegraphics[scale=0.5]{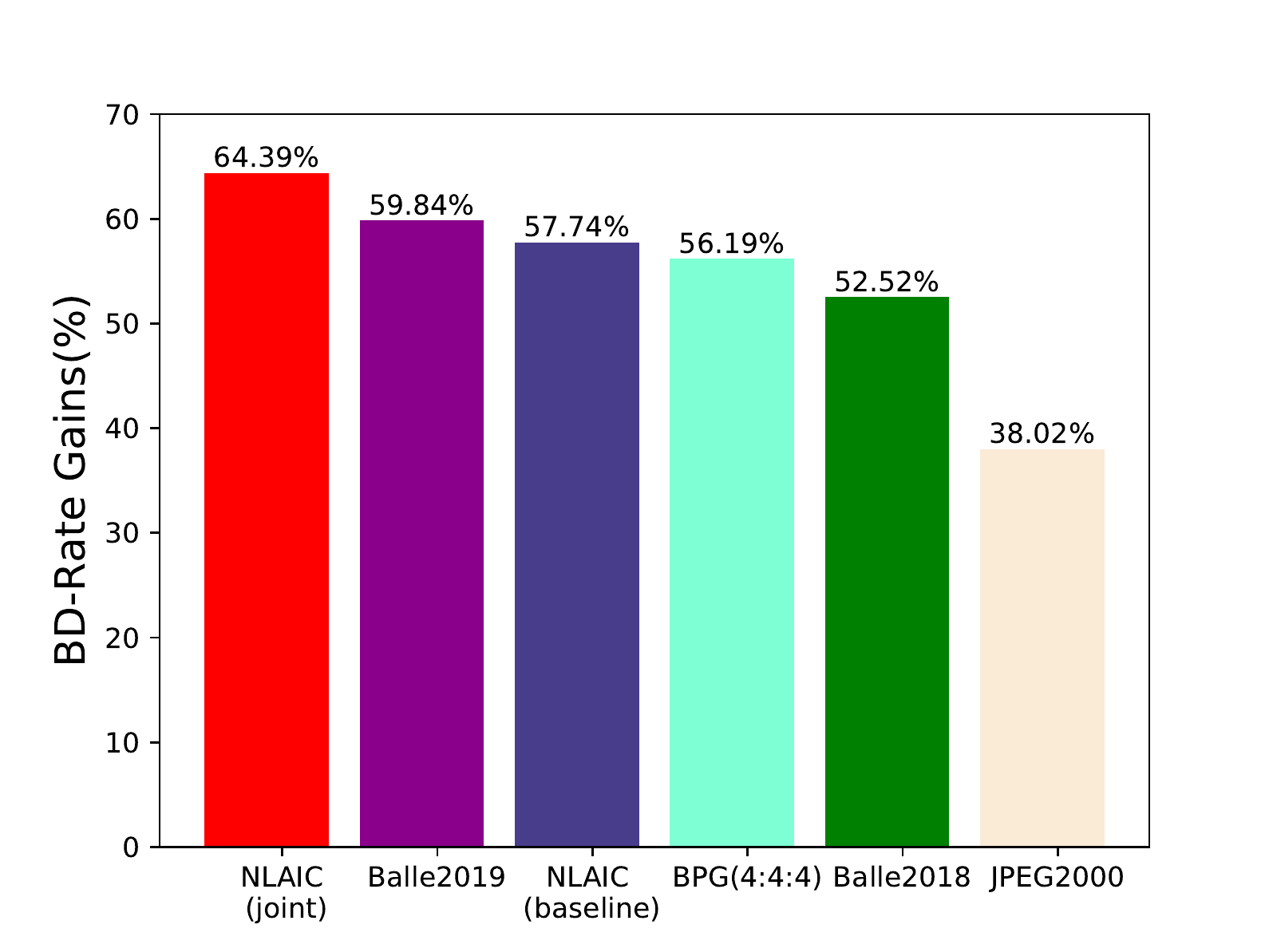}
   \caption{Coding efficiency comparison using JPEG as anchor. It shows our NLAIC achieves the best BD-Rate gains among all popular algorithms.}
   \label{bd_rate_reduction_bar}
\end{figure}
%For a model transfer from baseline model to the joint training model, we also fixed the baseline and train the context model alone to further remove the dependencies. It performs a little worse compared with the joint train model and only achieves around 3\% bit rate reduction with the same reconstruction.

\subsection{Performance Efficiency}
We evaluate our NLAIC models by comparing the rate-distortion performance averaged on publicly available Kodak dataset. Fig.~\ref{rd_curve} shows the performance when distortion is measured by MS-SSIM and PSNR, respectively, that are widely used in image and video compression tasks.
Here, PSNR represents the pixel-level distortion while MS-SSIM describes the structural similarity. MS-SSIM is reported to offer higher correlation with human perceptual inception, particularly at low bit rate~\cite{wang2003multiscale}. As we can see, our NLAIC provides the state-of-the-art performance with noticeable performance margin compared with the existing leading methods, such as Ball{\'e}2019~\cite{minnen2018joint} and Ball{\'e}2018~\cite{balle2018variational}.

Specifically, as shown in Fig.~\ref{sfig:ssim_perf} using MS-SSIM for both loss and final distortion measurement,  ``NLAIC baseline'' outperforms  the existing methods while the ``NLAIC joint'' presents even larger performance margin. For the case that uses MSE as loss and PSNR as distortion measurement, ``NLAIC joint'' still offers the best performance, as illustrated in Fig.~\ref{sfig:psnr_perf}.  ``NLAIC baseline'' is slightly worse than the model in~\cite{minnen2018joint} that uses contexts from both hyperpriors and neighbors jointly as our ``NLAIC joint'', but  better than the work~\cite{balle2018variational} that only uses the hyperpriors to do contexts modeling for a fair comparison. Fig.~\ref{bd_rate_reduction_bar} compares the average BD-Rate reductions by various methods over the legacy JPEG encoder. Our ``NLAIC joint'' model shows 64.39\% and 12.26\% BD-Rate~\cite{bjontegaard2001calculation}  reduction against JPEG420 and BPG444, respectively.

\subsection{Ablation Studies}
\begin{figure}[t]
   \centering
   \includegraphics[scale=0.3]{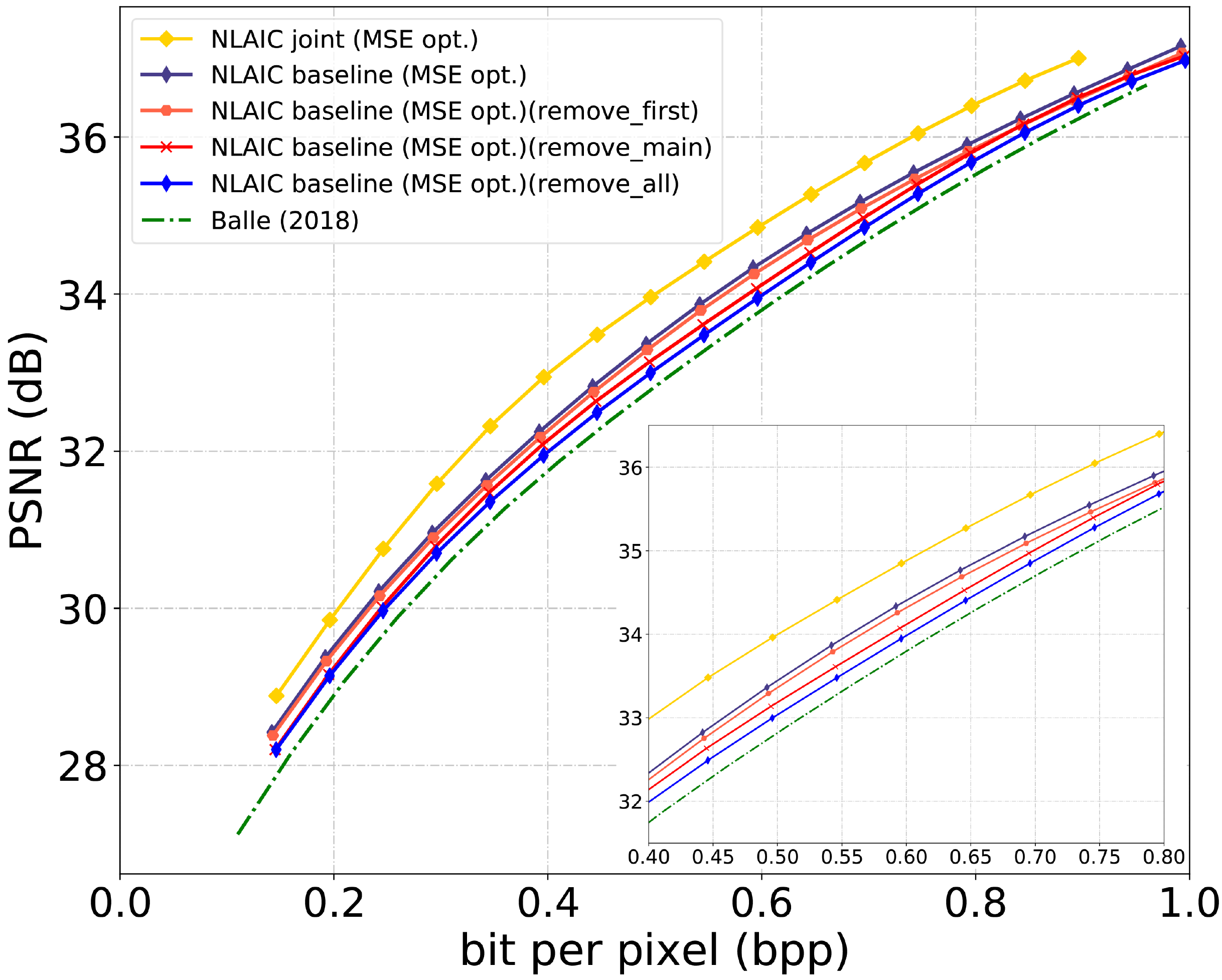}
   \caption{Ablation studies on NLAM where we gradually remove the NLAM components and re-train the model}
   \label{model_variants_discuss}
\end{figure}

We further analyze our NLAIC in following aspects:\\
{\bf Impacts of NLAM:} To further discuss the efficiency of newly introduced NLAM, we remove the mask branch in the NLAM pairs gradually, and retrain our framework for performance evaluation. For this study, we use the baseline context modeling in all cases, and use the MSE  as the loss function and PSNR as the final distortion measurement, shown in Fig.~\ref{model_variants_discuss}. For illustrative understanding, we also provide two anchors, i.e., ``Ball{\'e}2018''~\cite{balle2018variational} and ``NLAIC joint'' respectively. However, to see the degradation caused by gradually removing the mask branch in NLAMs, one should compare with the NLAIC baseline curve.

 Removing the mask branches of the first NLAM pair in the main encoder-decoders (referred to as ``remove\_first'') yields a PSNR drop of about 0.1dB compared to ``NLAIC baseline'' at the same bit rate. PSNR drop is further enlarged noticeably when removing all NLAM pairs' mask branches in main encoder-decoders (a.k.a., ``remove\_main''). It gives the worst performance when further disabling the NLAM pair's mask branches in hyperprior encoder-decoders, resulting in the traditional variational autoencoder without non-local characteristics explorations (i.e., ``remove\_all'').

 \begin{figure}[t]
   \centering
   \includegraphics[scale=0.18]{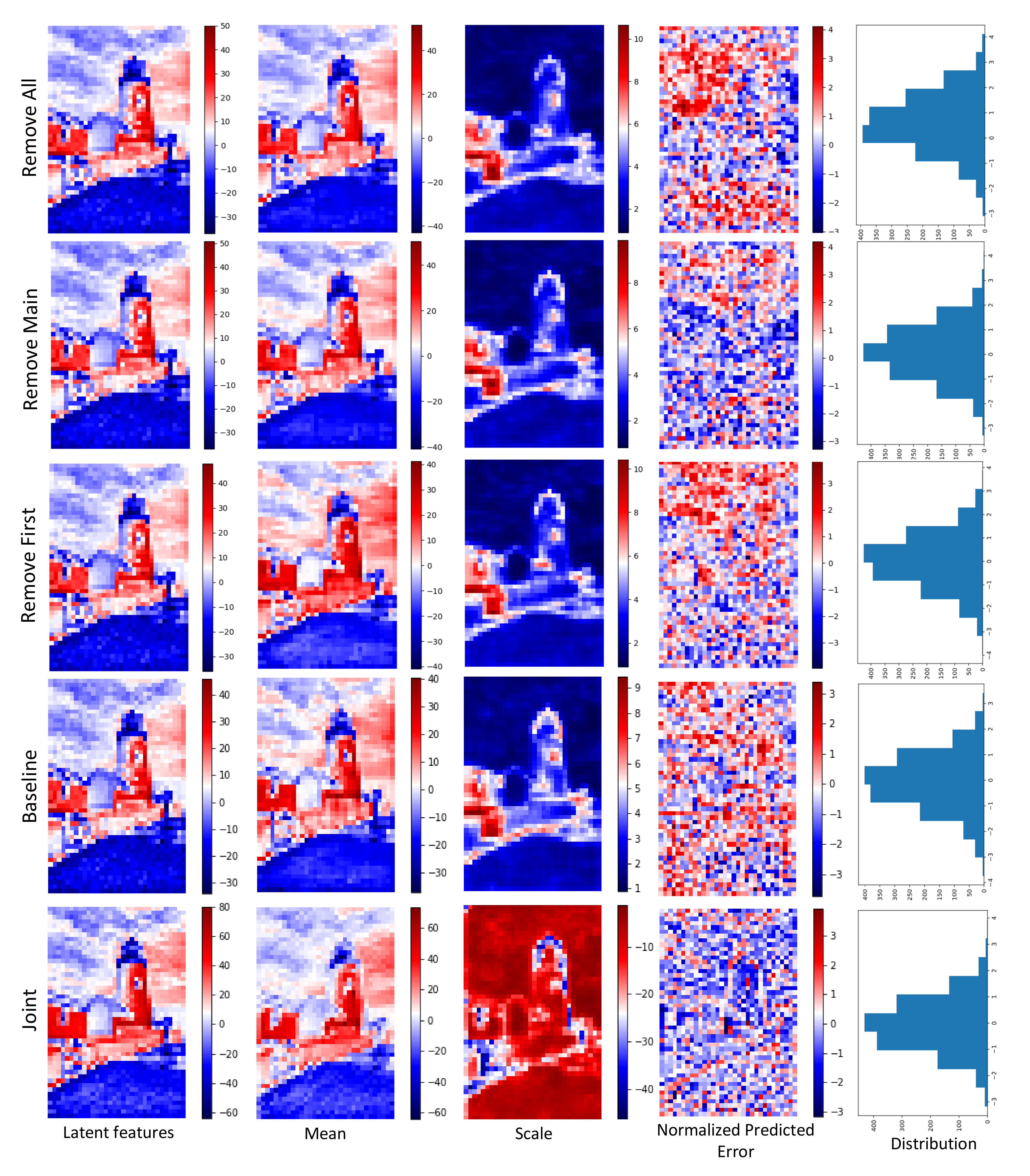}
   \caption{Prediction error with different model at similar bit rate. Column-wisely, it depicts the latent features, the predicted mean, predicted scale,  normalized prediction error ( i.e., $\frac{feature-mean}{scale}$) and the distribution of the normalized prediction error from left to right plots. Each row represents a different model (e.g., various combinations of NLAM components, and contexts prediction). These figures show that with NLAM and joint contexts from hyperprior and autoregressive neighbors, the latent features capture more information (indicated by a layer dynamic range), which leads to a large scale (standard deviation of features), and the final normalized feature prediction error has the most compact distribution, which leads to the lowest bit rate. }
   \label{prediction_error}
\end{figure}

{\bf Impacts of Joint Contexts Modeling:} We further compare  conditional context modeling efficiency of the model variants in Fig.~\ref{prediction_error}. As we can see,
with embedded NLAM and joint contexts modeling, our ``NLAIC joint'' could provide more powerful latent features,
and more compact normalized feature prediction error, both contributing to its leading coding efficiency.

{\textbf{Hyperpriors }$ \bf{\hat{z}:}$} Hyperpriors $\hat{z}$ has noticeable contribution to the overall compression performance ~\cite{minnen2018joint,balle2018variational}. Its percentage decreases as the overall bit rate increases, shown in Fig.~\ref{hyper_bits_comsuming}. The percentage of $\hat{z}$ for MSE loss optimized model is higher than the case using MS-SSIM loss optimization.
Another interesting observation is that $\hat{z}$ exhibits contradictive distributions of joint and baseline models, for respective MSE and MS-SSIM loss based schemes.  More explorations is highly desired in this aspect to understand the bit allocation of hyperpriors in our future study.

\begin{figure*}[t]
\centering
\includegraphics[scale=0.4]{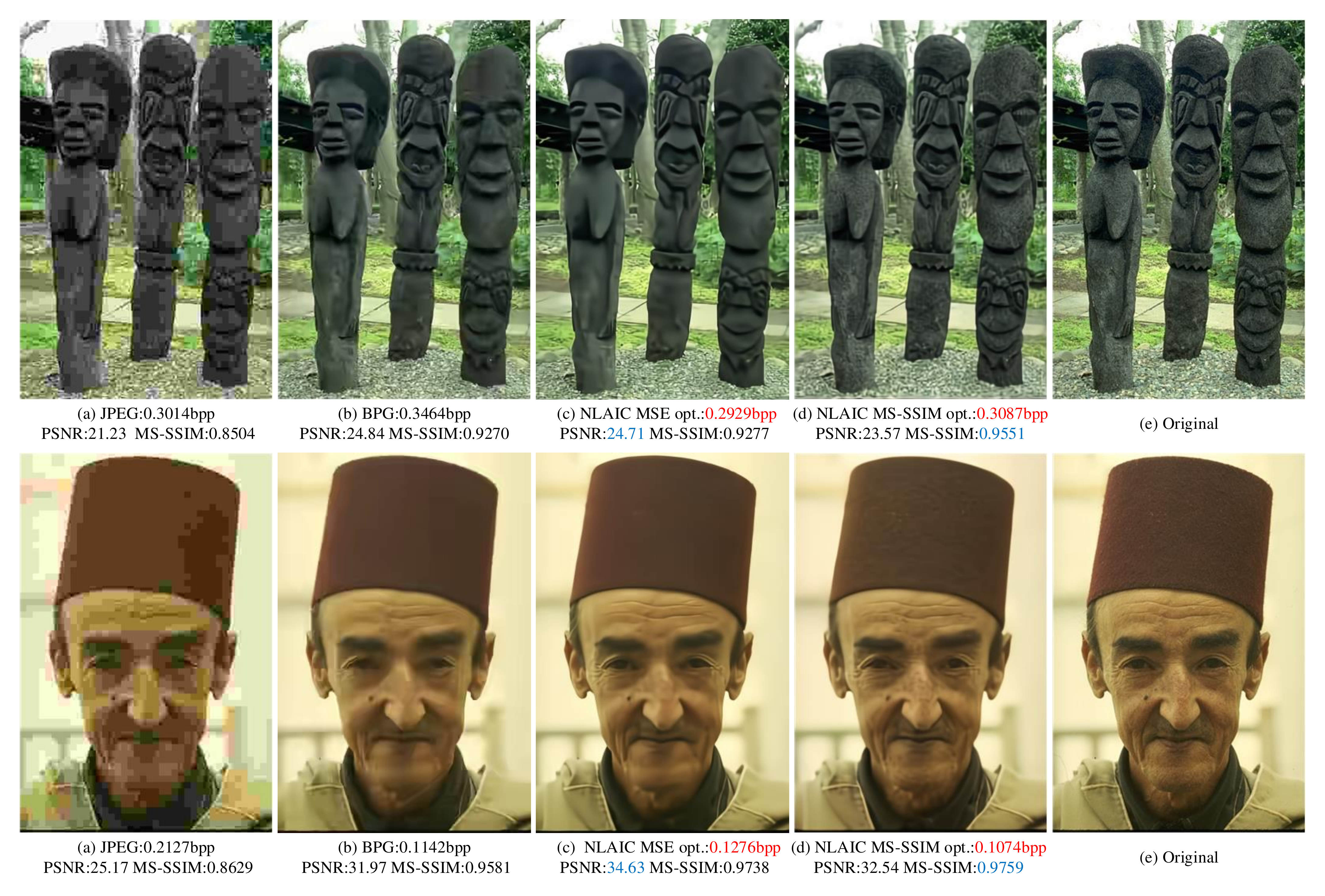}
%\subfigure[JPEG:0.3014bpp $\protect\\$  PSNR:21.23 $\protect\\$ MS-SSIM:0.8504]{\includegraphics[scale=0.305]{/1_1.png}}
%\subfigure[BPG:0.3464bpp $\protect\\$PSNR:24.84 MS-SSIM:0.9270]{\includegraphics[scale=0.305]{/1_2.png}}
%\subfigure[MSE opt:{\color{red}0.2929bpp}$\protect\\$ PSNR:24.71 MS-SSIM:0.9277]{\includegraphics[scale=0.305]{/1_PS.png}}
%\subfigure[MS-SSIM opt:{\color{red}0.3087bpp} $\protect\\$ PSNR:23.57 MS-SSIM:0.9551]{\includegraphics[scale=0.305]{/1_3.png}}
%\subfigure[Original]{\includegraphics[scale=0.305]{/1_4.png}}
%\subfigure[JPEG:0.2127bpp $\protect\\$  PSNR:25.17 MS-SSIM:0.8629]{\includegraphics[scale=0.30]{/2_1.png}}
%\subfigure[BPG:0.1142bpp  $\protect\\$  PSNR:31.97 MS-SSIM:0.9581]{\includegraphics[scale=0.30]{/2_2.png}}
%\subfigure[MSE opt:{\color{red}0.1276bpp} $\protect\\$ PSNR:34.63 MS-SSIM:0.9738]{\includegraphics[scale=0.30]{/35_PS.png}}
%\subfigure[MS-SSIM opt:{\color{red} 0.1074bpp } $\protect\\$  PSNR:32.54 MS-SSIM:0.9759]{\includegraphics[scale=0.30]{/2_3.png}}
%\subfigure[Original]{\includegraphics[scale=0.30]{/2_4.png}}
\caption{Visual comparison among JPEG420, BPG444, NLAIC joint MSE opt., MS-SSIM opt. and the original image from left to right. Our method achieves the best visual quality containing more texture without blocky nor blurring artifacts.
}
\label{visual_comparison}
\end{figure*}
%-------------------------------------------------------------------------
\begin{figure*}[t]
\centering
\includegraphics[scale=0.3]{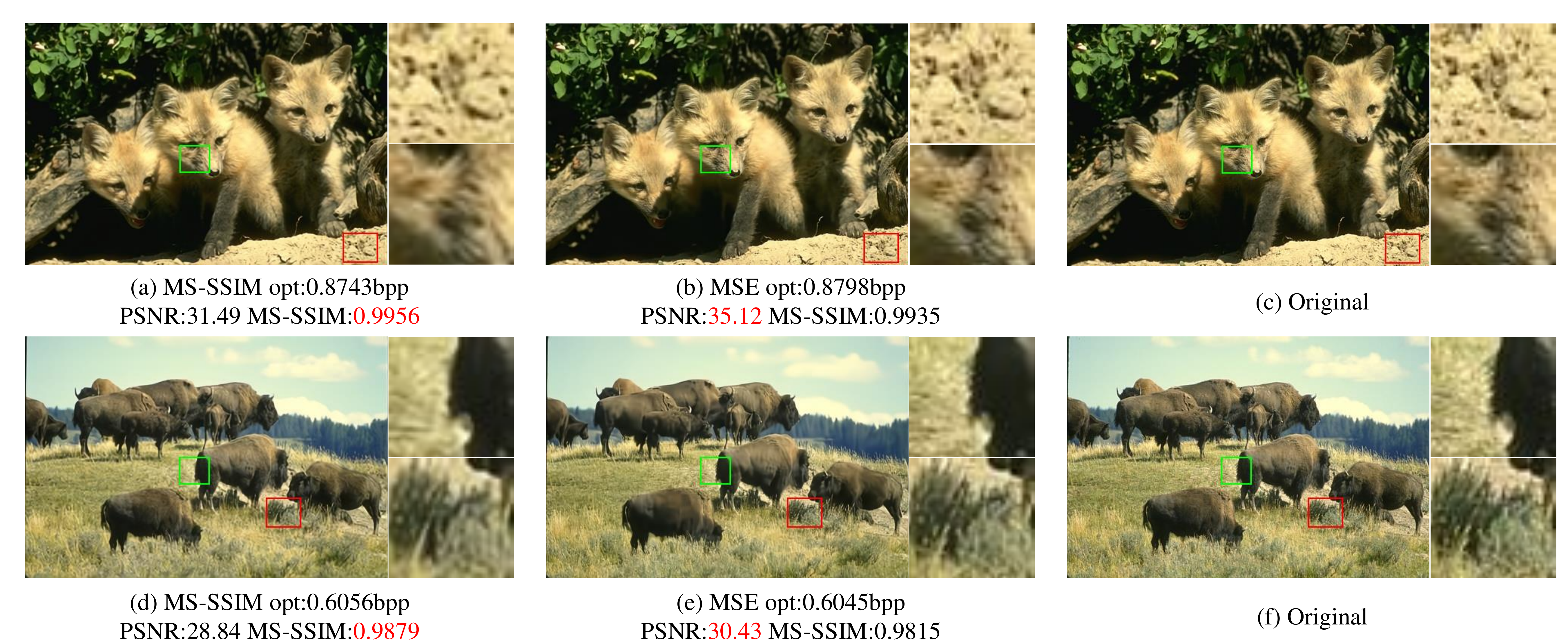}
\caption{Illustrative reconstruction samples of respective PSNR and MS-SSIM loss optimized compression}
\label{psnr_ssim_comparision}
\end{figure*}

\subsection{Visual Comparison}
We also evaluate our method on BSD500~\cite{bsd500} dataset, which is widely used in image restoration problems. Fig.~\ref{visual_comparison} shows the results of different image codecs at the similar bit rate. Our NLAIC provides the best subjective quality with relative smaller bit rate\footnote{In practice, some bit rate points cannot be reached for BPG and JPEG. Thus we choose the closest one to match our NLAIC bit rate.}.

Considering that MS-SSIM loss optimized results demonstrate much smaller PSNR at high bit rate in Fig.~\ref{sfig:ssim_perf},
we also show our model comparison optimized for respective PSNR and MS-SSIM loss at high bit rate scenario.
We find it that MS-SSIM loss optimized results exhibit worse details compared with PSNR loss optimized models at high bit rate, as shown in Fig.~\ref{psnr_ssim_comparision}.
This may be due to the fact that  pixel distortion becomes more significant at high bit rate, but structural similarity puts more weights at  a fair low bit rate. It will be interesting to explore  a better metric to cover the advantages of PSNR at high bit rate and MS-SSIM at low bit rate for an overall optimal efficiency.

%------------------------------------------------------------------------
\section{Conclusion}

In this paper, we proposed a non-local attention optimized deep image compression (NLAIC) method and achieve the state-of-the-art performance. Specifically, we have introduced the non-local operation to capture both local and global correlation for more compact latent feature representations. Together with the attention mechanism, we can enable the adaptive processing of latent features by allocating more bits to important area using the attention maps generated by non-local operations. Joint contexts from autoregressive spatial-channel neighbors and hyperpriors are leveraged to improve the entropy coding efficiency.

Our NLAIC outperforms the existing image compression methods, including well known BPG, JPEG2000, JPEG as well as the most recent learning based schemes~\cite{minnen2018joint,balle2018variational,rippel2017real}, in terms of both MS-SSIM and PSNR evaluation at the same bit rate.

For future study, we can make our context model deeper to improve image compression performance.
Parallelization and acceleration are important to deploy the model for actual usage in practice,
particularly for mobile platforms. In addition, it is also meaningful to extend our framework
for end-to-end video compression framework with more priors acquired from spatial and temporal information.

%{\small
\bibliographystyle{ieee}
\bibliography{non_local_image_compression_arxicfinal.bbl}
%}

\end{document}